\documentclass[letterpaper,aps,prd,twocolumn,preprintnumbers,showkeys,nofootinbib,nobibnotes]{revtex4}
\usepackage{epsfig}
\usepackage{dcolumn}
\usepackage{bm}

\makeatletter
\def\btt#1{\texttt{\@backslashchar#1}}%
\DeclareRobustCommand\bblash{\btt{\@backslashchar}}%
\makeatother

\begin{document}

\title{Constructing dark energy models with late time de Sitter attractor}

\author{Jian-gang Hao}
\author{Xin-zhou Li}\email{kychz@shtu.edu.cn}
\affiliation{SUCA, Shanghai United Center for Astrophysics,
Shanghai Normal University, 100 Guilin Road, Shanghai 200234,China
}%

\date{\today}

\begin{abstract}
ABSTRACT: In this paper, we describe a way to construct a class of
dark energy models that admit late time de Sitter attractor
solution. In the canonical scalar and Born-Infeld scalar dark
energy models, we show mathematically that a simple sufficient
condition for the existence of a late time de Sitter like
attractor solution is that the potentials of the scalar field have
non-vanishing minimum while this condition becomes that the
potentials have non-vanishing maximum for the phantom models.
These attractor solutions correspond to an equation of state
$w=-1$ and a cosmic density parameter $\Omega_{\phi}=1$, which are
important features for a dark energy model that can meet the
current observations.
\end{abstract}



\maketitle

\vspace{0.4cm} \noindent \textbf{I. Introduction} \vspace{0.4cm}

Current observations converge on that roughly two thirds of the
energy density in our universe is resulted from a kind of dark
energy that has negative pressure and can drive the accelerating
expansion of the universe\cite{riess,
perlmutter,bennett,Netterfield,Halverson}. Many candidates for
dark energy have been proposed so far to fit the current
observations. Among these models, the most typical ones are
cosmological constant and a time varying scalar field with
positive or negative kinetic energy evolving in a specific
potential, referred to as "quintessence"\cite{ratra,Coble,
Steinhardt,Peebles} or
"phantom"\cite{caldwell,Carroll,Frampton,hao1,li1,Gibbons,Feinstein,Sami}.
Successful dark energy models also share some common features: (i)
they should have an effective equation of state $w<-1/3$ so as to
accelerate the expansion of the universe at recent epoch. (ii)
they should be negligible compared with radiation and matter in
the early epoch of the universe so as not to affect the primordial
nucleosynthesis while dominate over the matter in a very recent
epoch. (iii) they should not be very sensitive to the initial
conditions so as to alleviate the fine tuning problems. Most
emphasis in the literatures is on the question of determining the
evolution of equation of state $w$. The purpose of this paper is
to clarify that the minimum (maximum for the phantom models) of
the potential corresponding to a cosmological de Sitter phase is a
dynamical attractor, to which a wide range of initial values will
converge.

The dynamical system of the scalar field with canonical Lagrangian
have been widely studied\cite{attractor}, among which the global
structure of the phase plane has been investigated and various
critical points and their physical significances have been
identified and manifested. However, in this paper, we focus on the
late time de Sitter attractor solution and give a very simple
sufficient condition for its existence. There are two major
motivations to study the dark energy models that have late time de
Sitter attractor. Firstly, a model will become very interesting if
its dynamical system admits a late time attractor solution that
leads to an equation of state $w=-1$ and a cosmic density
parameter $\Omega_{\phi}=1$, which meet all the above 3 points. In
this paper, we will show that if the potential of a model with
positive kinetic energy (quintessence) has non-vanishing minimum,
or the potential of a model with negative kinetic energy (phantom)
has non-vanishing maximum, the dynamical system of the model will
admit late time attractor solutions corresponding to $w=-1$ and
$\Omega_{\phi}=1$. The subsequent numerical study on some specific
models confirm the above conclusion. Secondly, recent observations
do not exclude, but actually suggest an equation of state
$-1.38<w<-0.82$\cite{Melchiorri}. A striking consequence of dark
energy with $w<-1$ is that the Universe will undergo a
catastrophic "big rip" in a finite time\cite{bigrip,McInnes}. If
dark energy is characterized by an equation of state $w<-1$
 (referred to as phantom in literatures) then the phantom energy
 density is still positive though it will first increase from a
 finite value up to infinite in a finite cosmic time, thereafter
 steadily decreases down to zero as cosmic time goes to infinity.
 To a fundamental observer in our galaxy, this state coincides
 with the above-mentioned big rip. However, in the case that there
 is a de Sitter attractor at late time, one can expect that the
 evolution of the scale factor recovers a rather conventional
 pattern, without big rip phase. de Sitter attractor will prevent
 the phantom energy density from increasing up to infinite in a finite
 cosmic time. Therefore, the presence of phantom energy does not
 lead to such a cosmic doomsday in a theory with de Sitter
 attractor at late time.

 The paper is organized as follows: In sections II and III, a sufficient
 condition for dark energy with late time de Sitter attractor for
 the canonical scalar field models without and with internal
 O(\textit{N}) symmetry was given. In section IV, the scalar field
 model in Born-Infeld type Lagrangian is discussed. In section V,
 the counterparts of the models in Section II, III and IV in the phantom
 scheme (with negative kinetic energy) have been investigated and the attractor behavior
 is manifested. In section VI, we numerically investigate some specific models and confirm
 the conclusion obtained in previous sections. Finally, in section
 VII, we present a discussion.

\vspace{0.4cm} \noindent\textbf{2. The scalar field model in
canonical Lagrangian}
 \vspace{0.4cm}

In this section, we will study the case that the dark energy is
mimicked by a scalar field expressed by canonical Lagrangian. We
will work in the flat Robertson-Walker metric
\begin{equation}\label{metric}
ds^{2}=-dt^{2}+a^{2}(t)(dx^2+dy^2+dz^2)
\end{equation}
The equation of motion for the scalar field with canonical
Lagrangian is

\begin{equation}\label{motion}
\ddot{\phi}+3H\dot{\phi}+\frac{\partial V(\phi)}{\partial \phi}=0
\end{equation}

To gain more insights into the dynamical system, we introduce the
new dimensionless variables

\begin{eqnarray}\label{newvarialbe1}
 x&&=\frac{\phi}{\phi_0}\nonumber\\
 y&&=\frac{\dot{\phi}}{\phi_0^2}\nonumber\\
 N&&=\ln a
\end{eqnarray}

\noindent Then the above equations could be rewritten as
\begin{eqnarray}\label{auto1}
 \frac{dx}{d N}&&=\frac{\phi_0 y}{H}\nonumber\\
 \frac{dy}{d N}&&=-3y-\frac{V'(x)}{\phi_0^3 H}
\end{eqnarray}

\noindent where the prime denotes the derivative with respect to
$x$ and $H$ is Hubble parameter that could be rewritten as
\begin{equation}\label{H1}
H^2=H_i^2E^2(N)
\end{equation}

\noindent where $H_i$ denote the Hubble parameter at an initial
time. $\Omega_{M, i}$ and $\Omega_{r, i}$ are the cosmic density
parameters for matter and radiation at the initial time. We also
choose the initial scale factor $a_i=1$. $E(N)$ is defined as
\begin{equation}\label{en}
E(N)=\left[\frac{\kappa}{3H_i^2}\left(\frac{\phi_0^4
y^2}{2}+V(x)\right)+\Omega_{M, i}e^{-3N}+\Omega_{r,
i}e^{-4N}\right]^{1/2}
\end{equation}

\noindent At late time, $N$ goes to be very large and the
contribution from matter and radiation in Eq.(\ref{H1}) become
negligible compared with the scalar field. To see this more
clearly, we take the example that when the equation of state of
the dark energy is constant and $w_{\phi}$ must be less than
$-1/3$ so as to accelerate the expansion of the universe. Thus the
dark energy component will evolve with $N$ as
$\Omega_{\phi}e^{-3(1+w_{\phi})N}$, which dissipates slower than
matter and radiation. So, at late time, we will have

\begin{eqnarray}\label{auto2}
 \frac{dx}{d N}&&=\sqrt{\frac{3}{\kappa}}\frac{\phi_0y}{\sqrt{\phi_0^4y^2/2+V(x)}}\nonumber\\
 \frac{dy}{d N}&&=-3y-\sqrt{\frac{3}{\kappa}}\frac{V'(x)}{\phi_0^3\sqrt{\phi_0^4y^2/2+V(x)}}
\end{eqnarray}

\noindent The critical point of the above autonomous system is
$(x_c, 0)$, where $x_c$ is defined by $V'(x_c)=0$. Linearize the
Eqs.(\ref{auto2}) about the critical point, we will have

\begin{eqnarray}\label{auto3}
 \frac{dx}{d N}&&=\sqrt{\frac{3}{\kappa V(x_c)}}\phi_0y\nonumber\\
 \frac{dy}{d N}&&=-\sqrt{\frac{3}{\kappa V(x_c)}}\frac{V''(x_c)x}{\phi_0^3}-3y
\end{eqnarray}

\noindent The eigenvalues of the system are

\begin{equation}\label{eigen}
\lambda_{1, 2}=\frac{-\alpha\pm\sqrt{\alpha^2-4\beta}}{2}
\end{equation}

\noindent where $\alpha=3$ and $\beta=\frac{3 V''(x_c)}{\kappa
V(x_c)\phi_0^2}$. Now, we can conclude that for a positive
potential $V(x)$, the critical point $(x_c, 0)$ is stable if
$V''(x_c)>0$. This is to say that the dynamical system has a
stable critical point at the minimum of the potential and this
critical point corresponds to a late time attractor solution.
Especially, from Eq.(\ref{eigen}), we can read that if
$\beta<\alpha$ the critical point will be a stable node and it
will be a stable spiral if $\beta>\alpha$. Although the relation
between $\alpha$ and $\beta$ will not alter the stability property
of the critical point, it will surely affect the way the field
approaching the attractor. If the critical point is a stable
spiral, the oscillation of the field will increase while for a
stable node, the field will approach the attractor rather
smoothly. These subtle properties have been further confirmed by
the numerical analysis in section VI. Next, let's read out the
physical implications when the system is at the attractor regime.
The cosmic density parameter for the dark energy is
\begin{equation}\label{densitypara1}
\Omega_{\phi}=\frac{\kappa[\phi_0^4y^2/2+V(x)]}{3H_i^2E^2(N)}
\end{equation}

\noindent and the equation of state of the scalar field is
\begin{equation}\label{eqofstate1}
w_{\phi}=\frac{\phi_0^4y^2-2V(x)}{\phi_0^4y^2+2V(x)}
\end{equation}
\noindent Clearly, from Eq.(\ref{densitypara1}) and
Eq.(\ref{eqofstate1}), one can find that $w_{\phi}=-1$ and
$\Omega_{\phi}=1$ at the late time attractor. Note that the energy
density of the scalar field at the critical point is $V(x_c)$ and
should not vanish, thus the sufficient condition for the existence
of a viable cosmological model with a late time de Sitter
attractor solution should be that: \emph{the potential of the
field has non-vanishing minimum.}

\vspace{0.4cm} \noindent\textbf{III. The scalar field with
internal O(\textit{N}) symmetry in canonical Lagrangian}
\vspace{0.4cm}

In this section, we devote ourselves to the scalar models in which
the fields possess an internal symmetry. The complex scalar (with
U(1) internal symmetry) field models were proposed by Gu and Huang
and Boyle \textit{et. al}.\cite{gu,Boyle}, and were generalized to
the general case with an O(\textit{N}) internal symmetry by Li,
Hao and Liu\cite{li}. The equation of motion for the O(\textit{N})
scalar field is
\begin{equation}\label{onmotion}
\ddot{\phi}+3H\dot{\phi}-\frac{\Sigma^2}{a^6\phi^3}+V'(\phi)=0
\end{equation}

\noindent where the term $-\frac{\Sigma^2}{a^6\phi^3}$ is resulted
from the internal symmetry and $\Sigma$ is an integration
constant\cite{li}. By using the dimensionless variables defined by
Eq.(\ref{newvarialbe1}), we can write the equation of motion as
\begin{eqnarray}\label{autoon1}
\frac{dx}{d N}&&=\frac{\phi_0y}{H}\nonumber\\
\frac{dy}{d N}&&=-3y-\frac{V'(x)}{\phi_0^3
H}+\frac{\Sigma^2e^{-6N}}{\phi_0^5x^3H}
\end{eqnarray}

\noindent where $H$ is given by Eq.(\ref{H1}) but with a different

\begin{eqnarray}\label{Eprime}
E(N)=&&\bigg[\frac{\kappa}{3H_i^2}\bigg(\frac{\phi_0^4y^2}{2}+
\frac{\Sigma^2e^{-6N}}{2\phi_0^2x^2}\nonumber\\&&+V(x)\bigg)+\Omega_{M,i
}e^{-3N}+\Omega_{r, i }e^{-4N}\bigg]^{1/2}
\end{eqnarray}

One can easily identify that the term
$\frac{\Sigma^2e^{-6N}}{2\phi_0^2x^2}$ from the contribution of
internal symmetry will decrease to be negligible at late time as
well as the terms from matter and radiation. Thus, the autonomous
system for the O(\textit{N}) scalar fields Eqs.(\ref{autoon1})
will reduce to Eqs.(\ref{auto2}) and the discussion in previous
section concerning the single scalar field still hold true here.
We will only write down the conclusion here: \emph{the
O(\textit{N}) scalar field models admit a late time attractor
solution when the potential has non-vanishing minimum
corresponding to $x_c\neq 0$}. This is because the equation of
motion Eqs.(\ref{onmotion}) will become singular at $x=0$ and
therefore $x_c\neq 0$ is a constraint imposed by the O(\textit{N})
symmetry. At the attractor regime, on can find that the cosmic
energy parameter for the O(\textit{N}) scalar fields is $
\Omega_{\phi}=1$ and the equation of state is $w_{\phi}=-1$, which
are the same as the single scalar field models except the process
it approach the late time attractor.

\vspace{0.4cm} \noindent\textbf{IV. The scalar field model in
Born-Infeld type Lagrangian}
 \vspace{0.4cm}

Recent work by Sen showed that the tachyon of string theory could
be described by Born-Infeld type Lagrangian and its roles in
inflation and dark energy have been studied\cite{sen, quint}. So,
it is also very interesting to consider the dark energy models
with Born-Infeld type Lagrangian. The equation of motion for a
scalar field with Born-Infeld type Lagrangian is
\begin{equation}\label{em2}
\ddot{T} +3H\dot{T}(1-\dot{T}^2)+\frac{V'(T)}{V(T)}(1-\dot{T}^2)=0
\end{equation}

By introducing the new dimensionless variables (Note that the
field in Born-Infeld Lagrangian has different dimension from that
in canonical Lagrangian)
\begin{eqnarray}\label{newvarialbe}
 x&&=\frac{T}{T_0}\nonumber\\
 y&&=\dot{T}
\end{eqnarray}

\noindent The equation of motion could be reduced to

\begin{eqnarray}\label{autonew1}
 \frac{dx}{d N}&&=\frac{y}{T_0 H}\nonumber\\
 \frac{dy}{d N}&&=-3y(1-y^2)-\frac{1-y^2}{T_0H}\frac{V'(x)}{V(x)}
\end{eqnarray}

\noindent where $H$ is the same as that defined by Eq.(\ref{H1})
but with a different

\begin{equation}\label{en2}
E(N)=\bigg[\frac{\kappa V(x)}{3H_i^2\sqrt{1-y^2}}+\Omega_{M,
i}e^{-3N}+\Omega_{r, i}e^{-4N}\bigg]^{1/2}
\end{equation}

In a similar fashion as in previous section, we conclude that the
contribution from matter and radiation to Hubble parameter become
negligible at late time. Thus Eqs.(\ref{autonew1}) becomes

\begin{eqnarray}\label{autonew2}
\frac{dx}{d
N}&&=\sqrt{\frac{3}{\kappa}}\frac{y}{T_0}\left[\frac{\sqrt{1-y^2}}{V(x)}\right]^{1/2}\nonumber\\
\frac{dy}{d
N}&&=-3y(1-y^2)-\sqrt{\frac{3}{\kappa}}\frac{V'(x)}{V(x)T_0}\left[\frac{\sqrt{1-y^2}}{V(x)}\right]^{1/2}(1-y^2)
\end{eqnarray}

\noindent The critical point is $(x_c, 0)$ and $V'(x_c)=0$.
Linearize the autonomous system about the critical point, we will
have
\begin{eqnarray}\label{autonew3}
\frac{dx}{d
N}&&=\sqrt{\frac{3}{\kappa V(x_c)}}\frac{y}{T_0}\nonumber\\
\frac{dy}{d N}&&=-\sqrt{\frac{3}{\kappa
V(x_c)}}\frac{V''(x_c)}{V(x_c)T_0}x-3y
\end{eqnarray}

\noindent One can observe that the linearized autonomous system
for the canonical scalar field are quite similar with that of the
Born-Infeld type scalar field except a factor of $1/V(x_c)$. Thus,
we can conclude that\emph{ for positive potentials, the
Born-Infeld type scalar model has a stable critical point if and
only if it has non-vanishing minimum, which correspond to the de
Sitter attractor solution of the system}. If the minimum of the
potential is zero, then the equation of motion of the field
Eqs.(\ref{em2}) will contain divergent term at the attractor. So,
for the scalar with Born-Infeld type Lagrangian, we have the same
sufficient condition for the existence of a viable cosmological
model for dark energy as well as for the scalar with canonical
Lagrangian.

\vspace{0.4cm} \noindent\textbf{V. The models with negative
kinetic energy----phantom}
 \vspace{0.4cm}

\noindent\textbf{A. Phantom model with canonical Lagrangian}
\vspace{0.4cm}

The phantom model with canonical Lagrangian has been widely
studied in literatures\cite{caldwell}. The equation of motion is
\begin{equation}\label{phantomeq}
\ddot{\phi}+3H\dot{\phi}-\frac{\partial V(\phi)}{\partial \phi}=0
\end{equation}

\noindent Similar to the discussion in section II, we can express
the equation systems of the phantom field in terms of the
dimensionless variables defined in Eq.(\ref{newvarialbe1}) as

\begin{eqnarray}\label{phantomauto1}
 \frac{dx}{d N}&&=\frac{\phi_0 y}{H}\nonumber\\
 \frac{dy}{d N}&&=-3y+\frac{V'(x)}{\phi_0^3 H}
\end{eqnarray}

\noindent where the prime denotes the derivative with respect to
$x$ and H is hubble parameter that could be rewritten as
\begin{equation}\label{phantomH1}
H^2=H_i^2E_{phan}^2(N)
\end{equation}

\noindent where $E_{phan}(N)$ is defined as
\begin{eqnarray}\label{phantomen}
E_{phan}(N)=&&\bigg[\frac{\kappa}{3H_i^2}\bigg(-\frac{\phi_0^4
y^2}{2}+V(x)\bigg)\nonumber\\&&+\Omega_{M, i}e^{-3N}+\Omega_{r,
i}e^{-4N}\bigg]^{1/2}
\end{eqnarray}

In a similar fashion as in section II, we can linearize
Eq.(\ref{phantomauto1}) about its critical point when the phantom
becomes dominant at late time.
\begin{eqnarray}\label{phantomauto3}
 \frac{dx}{d N}&&=\sqrt{\frac{3}{\kappa V(x_c)}}\phi_0y\nonumber\\
 \frac{dy}{d N}&&=\sqrt{\frac{3}{\kappa V(x_c)}}\frac{V''(x_c)x}{\phi_0^3}-3y
\end{eqnarray}

By comparing Eqs.(\ref{auto3}) and Eqs.(\ref{phantomauto3}), one
can easily obtain that for a positive potential, the system admits
stable critical points $(x_c, 0)$ when $V''(x_c)<0$. Therefore,
the sufficient condition for a viable cosmological phantom model
in canonical Lagrangian should be that the potential has
non-vanishing maximum. It is not difficult to observe that the
equation of state $w_{phan}=-1$ and the cosmic density parameter
$\Omega_{phan}=1$ at the critical point.

\vspace{0.4cm}

\noindent\textbf{B. O(\textit{N}) Phantom model with canonical
Lagrangian} \vspace{0.4cm}

In this subsection, we apply the above discussion to the
O(\textit{N}) scalar field with negative kinetic energy
(O(\textit{N}) phantom). The equation of motion for the
O(\textit{N}) phantom is
\begin{equation}\label{onphantomeq}
\ddot{\phi}+3H\dot{\phi}-\frac{\Sigma^2}{a^6\phi^3}-V'(\phi)=0
\end{equation}

By introducing the dimensionless variables as
Eq.(\ref{newvarialbe1}), we can write Eq.(\ref{onphantomeq}) as

\begin{eqnarray}\label{autoophantn1}
\frac{dx}{d N}&&=\frac{\phi_0y}{H}\nonumber\\
\frac{dy}{d N}&&=-3y+\frac{V'(x)}{\phi_0^3
H}+\frac{\Sigma^2e^{-6N}}{\phi_0^5x^3H}
\end{eqnarray}

\noindent where $H$ is given by Eq.(\ref{phantomH1}) but with a
different

\begin{eqnarray}\label{Eprime}
E_{phan}(N)=&&\bigg[\frac{\kappa}{3H_i^2}\bigg(-\frac{\phi_0^4y^2}{2}-
\frac{\Sigma^2e^{-6N}}{2\phi_0^2x^2}+V(x)\bigg)\nonumber\\&&+\Omega_{M,i
}e^{-3N}+\Omega_{r, i }e^{-4N}\bigg]^{1/2}
\end{eqnarray}

\noindent It is obvious that one can apply the same discussion as
in the subsection \textbf{V. A} to the O(\textit{N}) phantom and
obtain the same sufficient condition for the existence of a viable
O(\textit{N}) phantom cosmological model.

\vspace{0.4cm}

\noindent\textbf{C. Phantom model with Born-Infeld type
Lagrangian} \vspace{0.4cm}

In this subsection, we study the phantom model with Born-Infeld
type Lagrangian, which has been recently proposed in
Ref.\cite{hao2}. The equation of motion for the model is

\begin{equation}\label{emphan2}
\ddot{T} +3H\dot{T}(1+\dot{T}^2)-\frac{V'(T)}{V(T)}(1+\dot{T}^2)=0
\end{equation}

By introducing the dimensionless variables as
Eq.(\ref{newvarialbe}), the equation of motion could be reduced to

\begin{eqnarray}\label{phanautonew1}
 \frac{dx}{d N}&&=\frac{y}{T_0 H}\nonumber\\
 \frac{dy}{d N}&&=-3y(1+y^2)+\frac{1+y^2}{T_0H}\frac{V'(x)}{V(x)}
\end{eqnarray}

\noindent where $H$ is defined by Eq.(\ref{phantomH1}) but with a
different
\begin{equation}\label{phanen2}
E_{phan}(N)=\left[\frac{\kappa
V(x)}{3H_i^2\sqrt{1+y^2}}+\Omega_{M, i}e^{-3N}+\Omega_{r,
i}e^{-4N}\right]^{1/2}
\end{equation}

In a similar fashion as in section \textbf{IV}, we conclude that
the contribution from matter and radiation to Hubble parameter
become negligible at late time. Thus Eqs.(\ref{phanautonew1})
becomes

\begin{eqnarray}\label{phanautonew2}
\frac{dx}{d
N}&&=\sqrt{\frac{3}{\kappa}}\frac{y}{T_0}\left[\frac{\sqrt{1+y^2}}{V(x)}\right]^{1/2}\nonumber\\
\frac{dy}{d
N}&&=-3y(1+y^2)+\sqrt{\frac{3}{\kappa}}\frac{V'(x)}{V(x)T_0}\left[\frac{\sqrt{1+y^2}}{V(x)}\right]^{1/2}(1+y^2)
\end{eqnarray}

\noindent The critical point is $(x_c, 0)$ and $V'(x_c)=0$.
Linearize the autonomous system about the critical point, we will
have
\begin{eqnarray}\label{phanautonew3}
\frac{dx}{d
N}&&=\sqrt{\frac{3}{\kappa V(x_c)}}\frac{y}{T_0}\nonumber\\
\frac{dy}{d N}&&=\sqrt{\frac{3}{\kappa
V(x_c)}}\frac{V''(x_c)}{V(x_c)T_0}x-3y
\end{eqnarray}

By comparing Eqs.(\ref{autonew3}) and Eqs.(\ref{phanautonew3}), we
can conclude that the sufficient condition for the existence of a
viable cosmological model with late time de Sitter attractor for
the phantom model with Born-Infeld type Lagrangian is same as that
for the phantom with canonical Lagrangian and is that the
potentials should have non-vanishing maximum.

\noindent\textbf{VI. Some specific examples }
 \vspace{0.4cm}

In this section we will numerically investigate the models with
specific potentials and confirm the previous conclusions. Firstly,
as examples, we study the scalar field, O(\textit{N}) scalar field
and phantom field with canonical Lagrangian in the so called
Mexico hat potential
\begin{equation}\label{higgsp}
 V(\phi)=\frac{\lambda}{4}(\phi^2-\sigma_0^2)^2+V_0
\end{equation}

\noindent One can easily identify that the potential has maximum
at $\phi=0$ and minimum at $\phi=\sigma_0$. Therefore, according
to our previous conclusion, we can construct a phantom dark energy
model if the field moves toward its maximum while we can make a
conventional dark energy model if the field evolves toward the
minimum. Whether the field evolves to its maximum or minimum
depends on the attraction domain of the initial value of field.
One could not completely solve the fine-tuning problem by such a
way, but can alleviate it. That is, a rather wide range of initial
values will evolve to the same attractor. In Fig.1--Fig.9, we plot
the numerical results that confirm our previous conclusions. In
order to compare them, we choose the same dimensionless parameters
as $\frac{\kappa\phi_0^4}{3H_i^2}=0.3333$,
$\frac{\sigma_0^2}{H_i\phi_0}=1.0000$ and
$\frac{\Sigma^2}{H_i\phi_0^5}=1.0000$. The plots begins with the
equipartition epoch $\Omega_M=\Omega_r$. It must be pointed out
that this numerical analysis is merely used as confirmation of the
previous analytical conclusion. It is not specially designed to
meet the current observation. It would be straightforward to make
the models meet the observation by tuning the parameters
carefully, which is not the purpose of this current work.
\begin{figure}
\epsfig{file=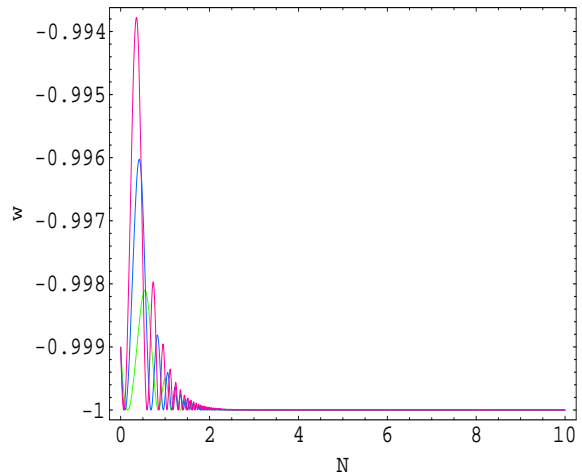,height=2.5in,width=3.0in} \caption{The
evolution of the equation of state of a real scalar field in
potential Eq.(\ref{higgsp}). The green, blue and pink curve
correspond to the choice of the dimensionless parameter
$\frac{\lambda\phi_0\sigma_0^2}{H_0}=2.0000, 4.0000, 6.0000$}
\end{figure}
\begin{figure}
\epsfig{file=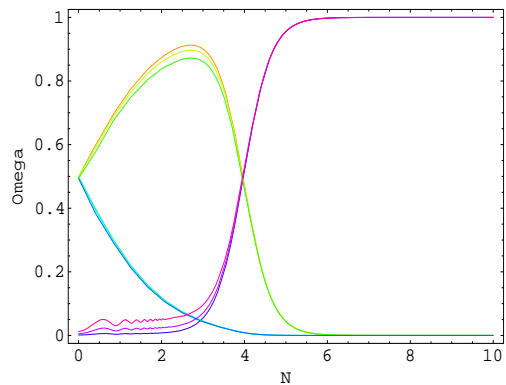,height=2.0in,width=2.6in} \caption{The
evolution of cosmic density parameter of matter $\Omega_M$(green,
yellow and orange curves), radiation $\Omega_r$ (blue curve) and
dark energy $\Omega_{\phi}$ (pink, red and indigo curve) in
potential Eq.(\ref{higgsp}). Since the parameters in the potential
will not affect the shape of the plot significantly, we here plot
for different initial $x$ and $y$, which affect the plot as shown
in the figure.}
\end{figure}
\begin{figure}
\epsfig{file=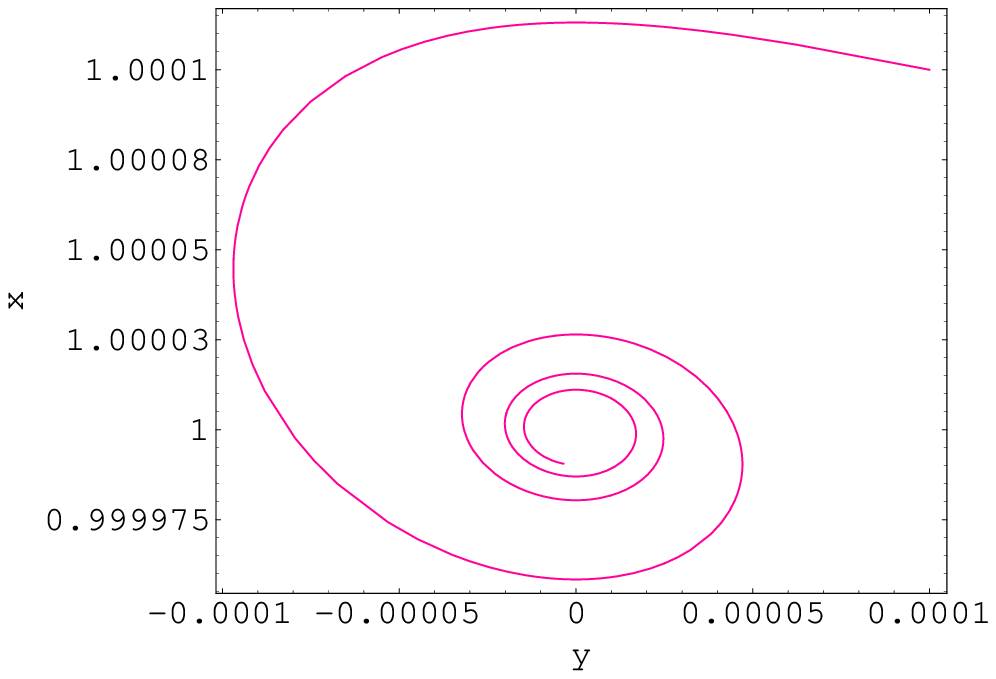,height=2.0in,width=3.0in} \caption{The
attractor property of the system in the phase plane for the real
scalar field model in potential Eq.(\ref{higgsp})}
\end{figure}

\begin{figure}
\epsfig{file=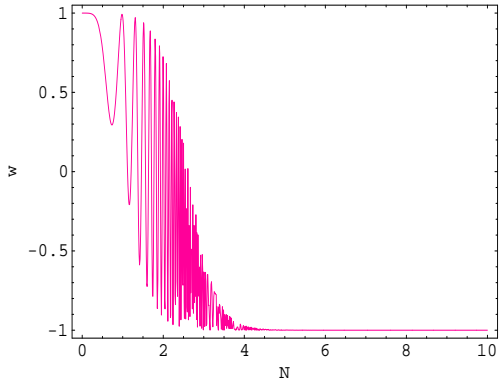,height=2.0in,width=2.6in} \caption{The
evolution of the equation of state of scalar fields with
O(\textit{N}) internal symmetry in potential Eq.(\ref{higgsp}).
The increase of oscillation is resulted from the introduction of
the internal symmetry.}
\end{figure}
\begin{figure}
\epsfig{file=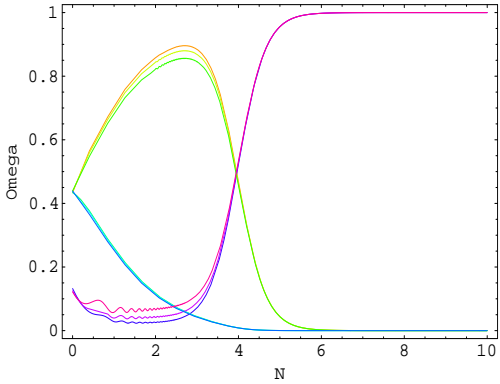,height=2.0in,width=2.6in} \caption{The
evolution of cosmic density parameter of matter $\Omega_M$(green,
yellow and orange curves), radiation $\Omega_r$ (blue curve) and
O(\textit{N}) scalar $\Omega_{\phi}$ (pink, red and indigo curve)
in potential Eq.(\ref{higgsp}). Note that the term
$\frac{\Sigma^2}{a^6\phi^3}$ resulted from the internal symmetry
make the evolution quite different from that without internal
symmetry at early epoch (the tilt of the yellow curve at the
initial state). Similar as in Fig.2, we here plot for different
initial $x$ and $y$, which affect the plot as shown in the figure.
}
\end{figure}
\begin{figure}
\epsfig{file=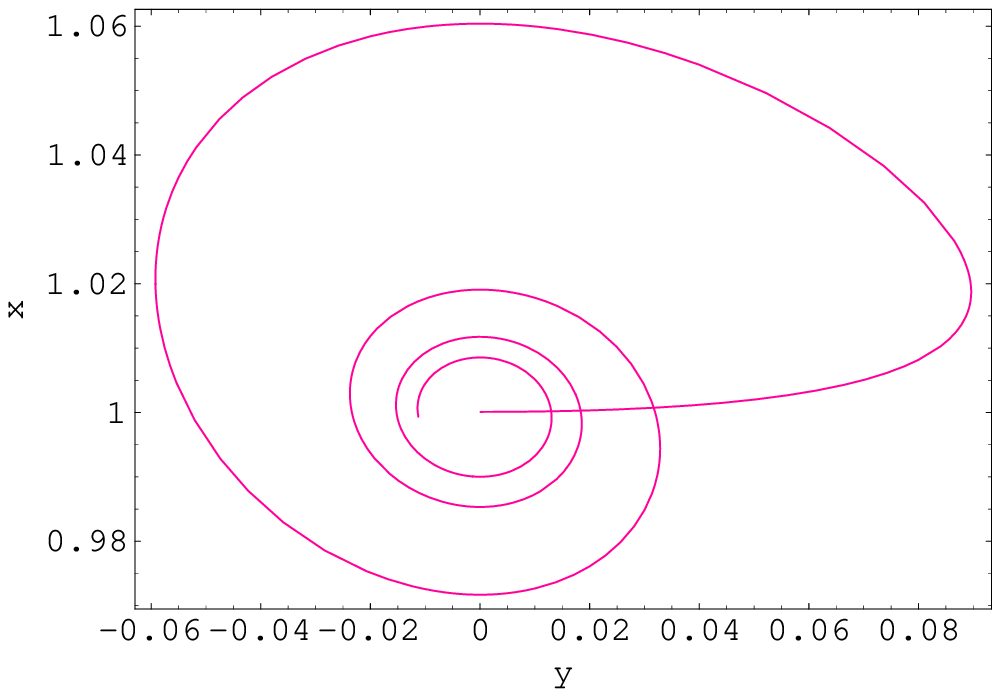,height=2.0in,width=2.6in} \caption{The
attractor property of the system in the phase plane for the
O(\textit{N}) scalar field model in potential Eq.(\ref{higgsp})}
\end{figure}

\begin{figure}
\epsfig{file=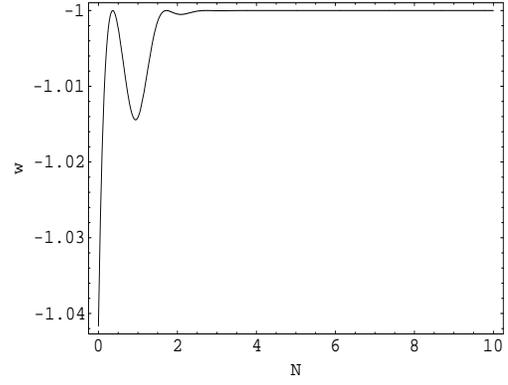,height=2.0in,width=2.6in} \caption{The
evolution of the equation of state of phantom field in potential
Eq.(\ref{higgsp})}
\end{figure}
\begin{figure}
\epsfig{file=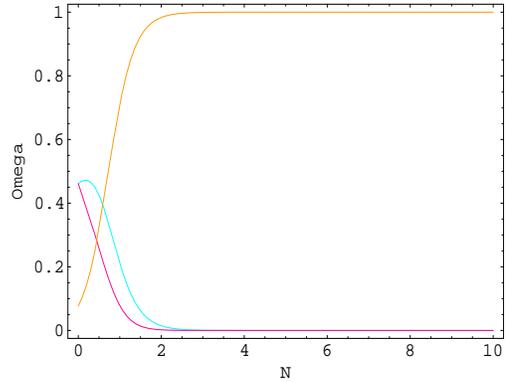,height=2.0in,width=2.6in} \caption{The
evolution of cosmic density parameter of matter $\Omega_M$(blue
curve), radiation $\Omega_r$ (red curve) and phantom
$\Omega_{\phi}$ (yellow curve) in potential Eq.(\ref{higgsp})}
\end{figure}
\begin{figure}
\epsfig{file=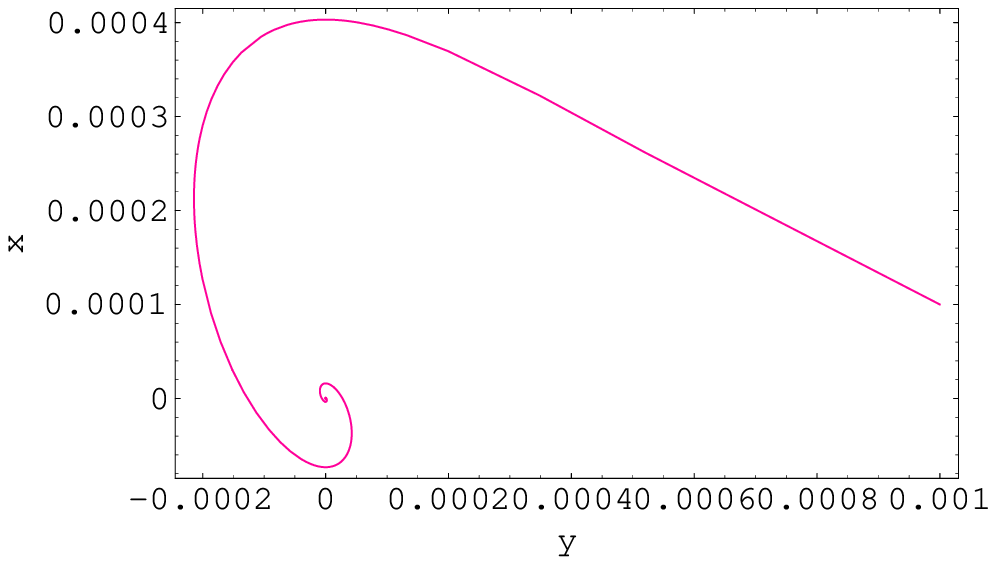,height=2.0in,width=2.6in} \caption{The
attractor property of the dynamical system in the phase plane for
the phantom model in potential Eq.(\ref{higgsp}) }
\end{figure}

One can also consider some other widely investigated potentials
such as the exponential potential $V(\phi)=V_0\exp(-\lambda
\kappa\phi)$ and inverse power potential
$V(\phi)=V_0(\phi_0/\phi)^q$. It easily observes that these
potentials admit vanishing minimum at $\phi\longrightarrow\infty$.
The field will evolve towards the minimum of the potentials while
can not reach them at finite time because the minimum corresponds
to the infinite $\phi$, and therefore the existence of late time
attractor solution for these types of potential need only the
existence of minimum and the "non-vanishing" condition is not
necessarily held.

 \vspace{0.4cm} \noindent\textbf{VII. Discussions}
 \vspace{0.4cm}

In this paper, we propose a way to construct viable dark energy
models featured by a late time de Sitter attractor. We show that
the dark energy model will admit a late time attractor solution if
the potential of the scalar field with both canonical and
Born-Infeld type Lagrangian has non-vanishing minimum. This
condition becomes that the potential has non-vanishing maximum for
the phantom models. When the field starts from the neighborhood of
the maximum (minimum for the phantom models), it will be attracted
to evolve towards the maximum/minimum, at which it will stay. This
stable attractor regime corresponds to the equation of state
$w_{\phi}=-1$ and cosmic density parameter $\Omega_{\phi}=1$,
which do not contradict with current observations. The current
observation data indicate that the cosmic density parameter of the
dark energy is about $\Omega_{\phi}=2/3$ and the equation of state
is less than $-0.82$\cite{Melchiorri}, therefore, in the models
analyzed in this paper, current universe is not at the attractor,
instead, it is just on the way to the attractor. It is necessary
to point out that the condition given in this paper for a viable
dark energy model is only a sufficient condition and thus strictly
speaking, it could not be used to rule out the possibility of the
existence of late time attractor solution for a given potential.
But it can tell us how to construct one.

\vspace{0.8cm} \noindent ACKNOWLEDGEMENT: This work was partially
supported by National Nature Science Foundation of China under
Grant No. 19875016 and Foundation of Shanghai Development for
Science and Technology under Grant No.01JC14035.

\end{document}